\documentclass[lettersize,journal]{IEEEtran}
\usepackage{amsmath,amsfonts}
\usepackage[caption=false,font=normalsize,labelfont=sf,textfont=sf]{subfig}
\usepackage{verbatim}
\usepackage{graphicx}

\usepackage[version=4]{mhchem} 
\usepackage{siunitx} 
\usepackage{hyperref} 
\usepackage{tikz} 
\usepackage{booktabs} 
\usepackage{mathtools} 
\usepackage{orcidlink} 
\usepackage{cprotect} 

\hypersetup{colorlinks,
    citecolor=black,
    linkcolor=black,
    urlcolor=black
}

\newcommand{\Lpath}{\mathcal{L}} 
\newcommand*{\defeq}{\stackrel{\text{def}}{=}} 
\newcommand*\diff{\mathop{}\!\mathrm{d}} 

\usepackage{subfiles} 

\begin{document}

\title{Reconstructed pCT Images Using Monte Carlo Simulations of a Scintillating Glass Detector}

\author{Adam Zieser \orcidlink{0000-0001-7618-3382}, Ugur Akgun \orcidlink{0000-0002-9850-4164}, Yasar Onel \orcidlink{0000-0002-8141-7769}
\thanks{Adam Zieser and Yasar Onel are with the Department of Physics \& Astronomy, University of Iowa, IA 52242 USA (e-mail: adam-zieser@uiowa.edu; yasar-onel@uiowa.edu)}%
\thanks{Ugur Akgun is with the Physics Department, Coe College, IA 52402 USA (e-mail: uakgun@coe.edu)}}%

\markboth{}%
{}
\maketitle

\begin{abstract}
The high cost and low image quality traditionally associated with proton computed tomography (pCT) have prevented it from seeing significant use in clinical settings. A cheap, compact, high-density scintillating glass detector capable of being attached to existing proton therapy gantries may help address these concerns. The design of the detector allows for use in conjunction with single-proton counting reconstruction algorithms, as well as beam-based algorithms that do not resolve individual protons within an accelerator bunch. This study presents quantitative reconstructed images of proton stopping power from Monte Carlo generated pCT scans using the radiation transport code MCNP6, demonstrating the feasibility of proton imaging using this detector design. Relative error and contrast have been examined and compared for images reconstructed using two reconstruction algorithms: a standard filtered backprojection algorithm to act as a benchmark, and a variant of a pCT algorithm which utilizes the concept of distance-driven binning.
\end{abstract}

\section{Introduction} \label{Intro}

\IEEEPARstart{P}{roton} therapy has become increasingly prevalent in the last two decades, with a recent increase in interest due in part to the possibility of utilizing the FLASH effect with ultra-high dose rates \cite{Esplen2020}. Protons deposit the majority of their energy in a sharp Bragg peak, followed by a sudden distal falloff, making them particularly attractive for use in treating cancers in sensitive or deep-seated areas, such as head and neck \cite{Moreno2019, Patel2014, Cozzi2001} or prostate cancers \cite{Slater2004}. Central to effective treatment planning is an accurate map of patient stopping power (SP), which is usually derived via x-ray CT stoichiometric calibration curve \cite{Schneider1996, Schaffner1998}. However, it is well-known that limitations in such CT-SP conversions introduce considerable SP range errors \cite{Knopf2013}, with studies reporting an average of anywhere from \SI{2.5}{\percent} \cite{Meinsdottir2019} to \SI{3.5}{\percent} \cite{Moyers2001, Moyers2010, Paganetti2012, Yang2012}, in addition to an absolute range error of \SI{1}{\mm} to \SI{7}{\mm} \cite{Meinsdottir2019, Taasti2019}. Proton CT (pCT) imaging \cite{Cormack1963, Cormack1976} has been proposed as an alternative to the stoichiometric calibration, using protons to directly image patient stopping power. Implementation of this imaging modality is made difficult due to the probabilistic nature of charged particle transport in matter, but the increased prevalence of proton therapy and recent advances in detector technology have renewed interest in the field---see Poludniowski et al. \cite{Poludniowski2015} for a comprehensive history.
%
\begin{figure}[t]
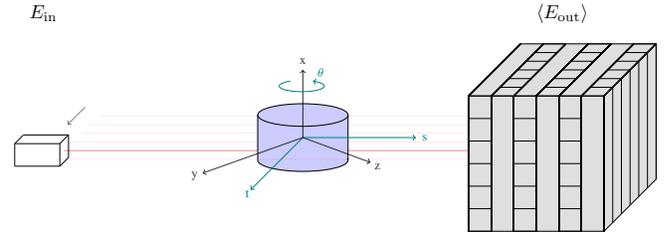

\centering
     \scalebox{0.6}{\subfile{anc/DetectorDiagram.tex}}  

\caption{Schematic diagram of the pCT setup, with proton source, phantom, and glass detector. Only five of the detector's \si{70} layers are shown, to demonstrate the alternating bar configuration.
\label{fig:DetectorDiagram}}
\end{figure}
%

In addition to problems of technical feasibility due to multiple Coulomb scattering, adoption of pCT has been impeded by the high costs associated with generating \SI{}{\MeV}-grade protons. Akgun et al. \cite{Wilkinson2017, Wilkinson2017_2, Varney2019} have proposed a cheap, compact, high-density scintillating glass calorimeter, which can be attached to existing proton therapy gantries for use in pCT or range verification in proton therapy treatment planning. The glass, which can be created with cheap reagents using a standard melt-quench glass-making technique, avoids the high costs associated with expensive scintillating crystals like LSO or BGO. The glass \cite{Tillman2017} has a density suitable for stopping \SI{}{\MeV}-scale protons, and is capable of resolving proton Bragg peaks up to a beam energy of \SI{175}{\MeV} within \SI{7}{\cm} of glass. Additionally, the geometry of the detector allows for the omission of one of the tracking planes used in typical pCT configurations, such as \cite{Schulte2004}.
%
\begin{table*}[bt]
\caption{Physical properties and elemental compositions (by weight) of the Gammex RMI 467 phantom. A general formula for grade 2 titanium alloy was chosen, and all other compositions have been taken from \cite{VanAbbema2015}.
\label{tab:GammexTable}}
\subfile{anc/PhantomTable.tex}
\end{table*}
%

To satisfy the detector's design specifications, pCT algorithms utilized with it should accommodate the particle rates associated with proton therapy pencil beams (PB). However, it has been noted \cite{Rescigno2015} that the particle rates of proton therapy accelerators may be too fast to make single-proton detection feasible. For example, the Varian ProBeam isochronous cyclotron (Varian Medical Systems, Palo Alto, California), which has a cyclotron frequency of \SI{72.8}{\MHz} \cite{Jolly2020}, will deliver an approximately \SI{1}{\ns}-bunch of about \si{86} protons per \SI{13.74}{\ns} RF cycle while operating at an average current of \SI{1}{\nA}. Regardless, the geometry of the detector is suitable for tracking individual protons or mean-beam paths, and utilizing either algebraic reconstruction (ART) techniques \cite{Li2006, Schulte2008, Wang2010, Penfold2010, CollinsFekete2017}, filtered backprojection (FBP) algorithms, or maximum-likelihood procedures \cite{CollinsFekete2016, Lazos2020}, making it versatile and adaptable to any situation. Accordingly, the base glass can be doped with millisecond decay time $ \ce{Eu^{3+}} $ ions for sufficiently slow beam scanning times, or nanosecond decay time $ \ce{Ce^{3+}} $ ions for single-proton ART and maximum-likelihood methods. Machine learning reconstruction techniques have also also been tested using this detector design \cite{Varney2019, Ademoski2019}.
%
\begin{figure}[t]
\centering
\includegraphics[width=0.45\textwidth]{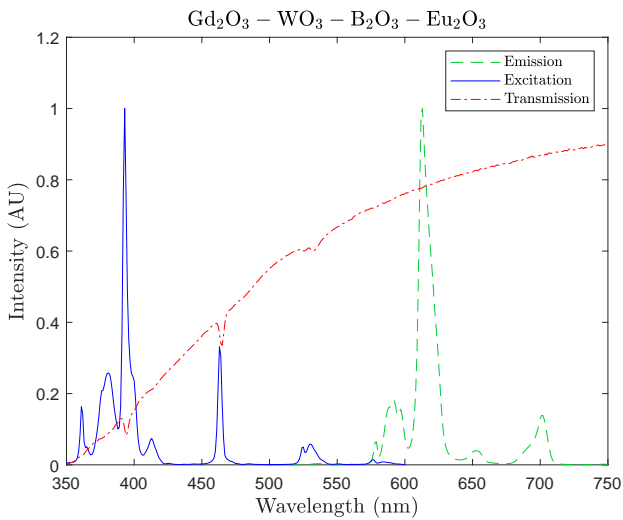}
\caption{Transmission, excitation, and emission spectra of the scintillating $ \ce{Gd2O3-WO3-B2O3-Eu2O3} $ glass, from Tillman et al. \cite{Tillman2017}. Curves have been normalized to arbitrary units with a max intensity of \si{1}, in the interest of overlaying all curves.
\label{fig:GlassSpectra}}
\end{figure}
%

In this study, Monte Carlo simulations of an idealized pCT scan of a commonly used calibration phantom were carried out, and the first quantitative stopping power image reconstructions on the proposed detector were performed, utilizing a reconstruction algorithm specifically designed for use with proton radiotherapy beams. FBP-style algorithms have been investigated as a fast alternative to algebraic reconstruction techniques in pCT \cite{Rit2013, Rescigno2015, Poludniowski2014, Krah2020, Khellaf2020, Dedes2022}, and are particularly suitable for reconstruction algorithms that track the collective behavior of proton bunches. Proton CT scans were simulated within the Monte Carlo radiation transport code, MCNP (version 6.1.1) \cite{MCNP6ReleaseNotes, MCNP61}. An FBP reconstruction utilizing distance-driven binning (DDB) \cite{Rit2013} was adapted from the work of Rescigno et al. \cite{Rescigno2015}, and uses the proton imaging equation of Wang et al. \cite{Wang2010}. Analyzing reconstructed SP images, contrast, and relative errors for each phantom insert, it is shown that this algorithm performs incrementally better than the standard FBP at select beam energies, and demonstrates the feasibility of the detector design.
\section{Materials and Methods} \label{MatsMethods}
\subsection{Scintillating glass properties} \label{GlassProperties}

The pCT images simulated in this study serve as a continuation of the proposal outlined by Akgun et al. \cite{Wilkinson2017, Wilkinson2017_2, Varney2019, Ademoski2019}, and are performed using simulations of the detector material in the compact glass proton imager. The detector is composed of \si{70} layers of $ \SI{100}{\mm} \times \SI{1}{\mm} \times \SI{1}{\mm} $ glass bars, arranged in an alternating orientation of \si{100} bars per layer, for a total size of $ \SI{10}{\cm} \times \SI{10}{\cm} \times \SI{7}{\cm} $. A visual representation of this arrangement is given in Fig. \ref{fig:DetectorDiagram}. In a completed prototype detector, silicon photomultipliers will be attached to the ends of each bar for data readout. However, only a stack of \si{70} glass plates ($ \SI{10}{\cm} \times \SI{10}{\cm} $ each) are simulated in this study, as only the average energy of a PB core exiting the phantom is required for the reconstruction algorithm employed, based on the method presented in Sec. \ref{ProtonBeams} and Sec. \ref{Reconstruction}. To utilize single-proton style algorithms, coordinate data (including position and incident angle) would be extracted from the individual bars by assuming each coordinate remains relatively unchanged over the distance of two layers (or \SI{2}{\mm}).

The simulated scintillating glass was based on previous work by Tillman et al. \cite{Tillman2017}, which determined that the optical and physical properties of a number of high-density glasses were suitable for proton imaging applications. Chosen for this work was a $ \ce{Gd2O3-WO3-B2O3} $ glass with a \SI{1}{\percent} europium-oxide dopant, in the system originally studied by Taki et al. \cite{Taki2013}. The optical and physical properties of the glass used in this study were acquired from samples manufactured with the base composition $ \ce{0.99(0.25 Gd2O3 + 0.55 WO3 + 0.2(2H3BO3))) + 0.01 Eu2O3} $ (by \SI{}{mol \percent}). All materials were of reagent grade quality, equal to or greater than \SI{99}{\percent} purity. For purposes of Monte Carlo modeling, it is assumed that the boric acid in the base composition has completely undergone the reaction $ \ce{2H3BO3 -> B2O3 + 3H2O} $, such that all hydrogen atoms have left the resultant glass. Most noteworthy is its density of \SI{5.84}{\g\per\cm\cubed}, which allows it to completely stop proton beams of nearly \SI{180}{\MeV} within the detector depth of \SI{7}{\cm}. Relevant optical properties are shown in Fig. \ref{fig:GlassSpectra}, demonstrating that the glass is transparent to its own emission spectrum and a large portion of its excitation spectrum, making it acceptable for use in a scintillation detector.
\subsection{pCT simulation in MCNP6} \label{Simulations}

All pCT simulations were carried out using MCNP6.1.1. Proton physics and parameters related to Monte Carlo transport are set to default MCNP parameters, with all secondary particles deposited locally. The latter assumption is dosimetrically incorrect, but no change in Bragg peak placement is seen compared to a full treatment including neutrons, electrons, and photons \cite{Lee2015}. Theoretically, the Bragg peak location is characteristic of a PB core, and is largely unaffected by the secondaries of the beam's so-called halo and aura \cite{Gottschalk2015}.

Scans with PB energies of \SI{150}{\MeV}, \SI{170}{\MeV}, \SI{190}{\MeV}, and \SI{210}{\MeV} were simulated, and proton histories were sampled from a one-dimensional spatial gaussian distribution, with no angular spread or beam emittance. The spatial spread of the beam is truncated by a cookie-cutter cell beyond two standard deviations. Two different beam spreads have been simulated: $ \sigma_t(0) = \SI{2.5}{\mm} $ and $ \sigma_t(0) = \SI{1.5}{\mm} $, with the former based on the specifications of the IBA (Louvain-la-Neuve, Belgium) dedicated nozzle \cite{Moignier2016} and the latter representing a more optimistic width similar to the beam spreads available at the Northwestern Medicine Chicago Proton Center \cite{Dedes2022}. To limit computation time, a \SI{1}{\mm} thick slice of the pCT universe was simulated (with all proton histories leaving this volume terminated), resulting in an approximately two-dimensional scan. To take advantage of reconstruction algorithms using parallel-beam geometry, \si{180} projections were taken over \SI{360}{\degree} (with \SI{2}{\degree} angle increments) and \si{141} beam rays per projection (with \SI{1}{\mm} spacing). These rays consisted of \si{1e4} proton histories, which were sufficient to reliably bring the coefficient of variation of all MCNP tallies below \SI{0.01}{}. Utilizing a number of PB bunches that provide this many protons may be more data than is required for pCT \cite{Sadrozinski2012}, but is sufficient to ensure convergent tally statistics within the PB core.

\begin{figure}[t]
\centering
\includegraphics[width=0.45\textwidth]{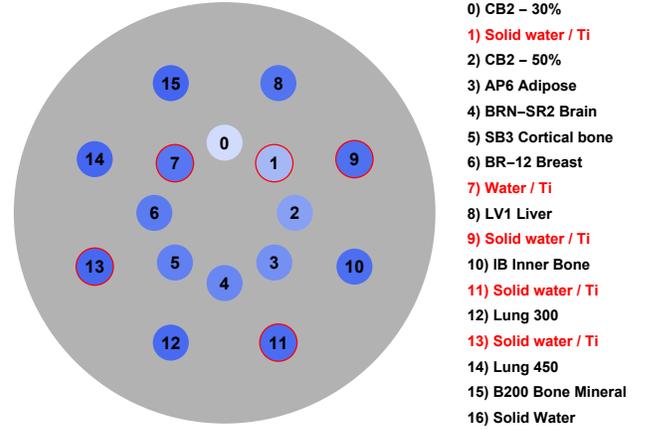}
\caption{Geometry and inserts of one 2D slice of the Gammex RMI 467 tissue calibration phantom in both simulated configurations. Inserts 1, 7, 9, 11, and 13 are highlighted, and change depending on the insert configuration. 
\label{fig:Gammex}}
\end{figure}

The phantom simulated was the Gammex RMI 467 tissue calibration phantom (Gammex Inc, Middleton, WI), with the disk scaled to one-third size, and the inserts to \SI{0.4}{}-scale. Two different insert configurations were explored to test the effects of extreme beam hardening: one ``standard" configuration suggested by the user's manual, and one ``titanium configuration" with all water-like inserts replaced with grade 2 titanium alloy. An illustrative diagram is given in Fig. \ref{fig:Gammex}, and a table of relevant physical characteristics is listed in Tab. \ref{tab:GammexTable}. In total, four full simulations were carried out at \SI{210}{\MeV}: one for each beam width and detector insert configuration. Due to the poor performance of the \SI{2.5}{\cm} beam, only two scans were simulated for the remaining three tested energies: one for each phantom configuration.
%
%

\begin{figure}[t]
\centering
\includegraphics[width=0.45\textwidth]{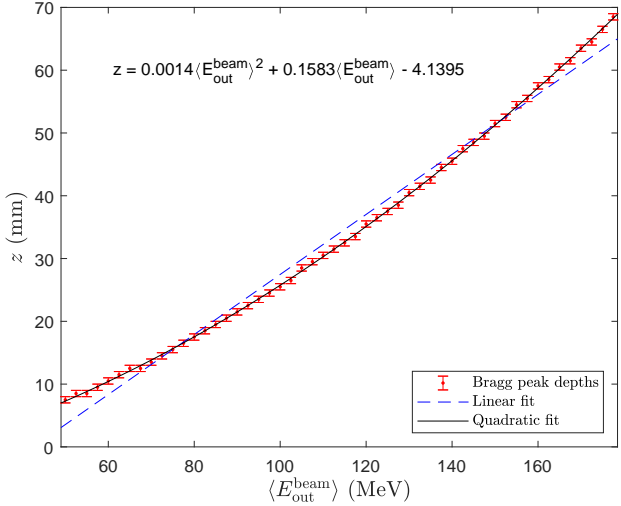}
\caption{Calibration curve of Bragg peak depth as a function of average beam energy incident upon the glass detector. Error bars of \SI{1}{\mm} correspond to the thickness of the glass bars, which limit the resolution of the Bragg peak. A linear fit is given for visual effect, to contrast with the relatively gently-sloping quadratic fit.
\label{fig:CalibrationCurve}}
\end{figure}

To avoid simulating the glass detector with every PB ray, the average energy of each beam was recorded after traversing the phantom, by dividing a \verb|*F1| total energy surface flux tally by an \verb|F1| particle flux surface tally. To convert these quantities into more realistic measurements, a quadratic calibration curve between Bragg peak depth and average beam energy was formed using a series of simulations with $ \langle E_\mathrm{out} \rangle $ ranging from \SI{50}{\MeV} to \SI{180}{\MeV} every \SI{2.5}{\MeV} (the results of which are seen in Fig. \ref{fig:CalibrationCurve}). This method is described in detail in Sec. \ref{Images}. Energy deposition, rather than scintillation light yield, is an acceptable quantity to tally in this context, as the two quantities are generally linearly related for non-organic scintillators, such that either one will record identical Bragg peak placement. 
\subsection{Proton beam characterization} \label{ProtonBeams}

The stopping power at proton therapy energies is given by the Bethe-Bloch equation (with no density or shell corrections) \cite{Newhauser2015} by
%
\begin{align}
\begin{split}
S(\rho_e, I, E) & = 4 \pi r_e^2 m_e c^2 \frac{\rho_e}{\beta(E)^2} \times \\ & \bigg[ \ln \bigg( \frac{2 m_e c^2}{I} \frac{\beta(E)^2}{1-\beta(E)^2} \bigg) - \beta(E)^2 \bigg].
\end{split}
\label{eq:Bethe}
\end{align}
%
In this study, the large-angle scatters of the halo are neglected, so the simulated PB fluence can be modeled as a gaussian packet according to Fermi-Eyges theory \cite{Gottschalk2012},
%
\begin{align}
\Phi(t, \mu_t, s) = N(s) \frac{\exp \Big(- \frac{1}{2} \frac{(t - \mu_t)^2}{\sigma_t^2 (s)} \Big) }{\sqrt{2 \pi} \sigma_t (s)},
\label{eq:Fluence}
\end{align}
%
with the coordinates $ (t, s) $ (shown in Fig. \ref{fig:DetectorDiagram}) being a standard Euclidean coordinate system $ (y, z) $ rotated by the angle $ \theta $, given by $ t = y \cos \theta + z \sin \theta $ and $ s = - y \sin \theta + z \cos \theta $. The beam is parameterized by a constant mean $ \mu_t $, and surviving proton primaries $ N(s) $ (described using a nuclear attenuation function \cite{Ulmer2007}). The depth-dependent variance $ \sigma_t^2 (s) $ is described using the differential scattering power $ T\big(pv(s)\big) $,
%
\begin{align}
\sigma^2_t (s) = \sigma^2_t (0) + \frac{1}{2} \int_0^s (s - s')^2 T\big(pv(s')\big) \diff s',
\label{eq:Sigma}
\end{align}
%
%
%
%
\begin{figure}[t]
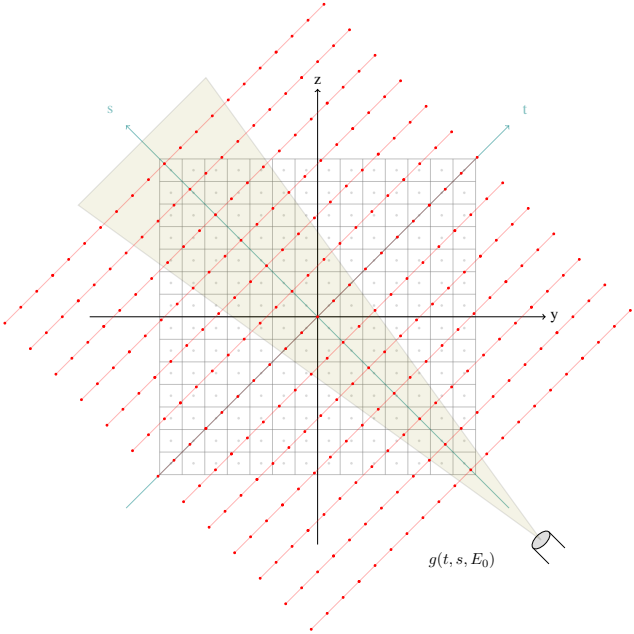

\centering
     \scalebox{0.6}{\subfile{anc/DDB_Grid.tex}}  

\caption{A diagram of the grid of virtual radiographs $ g(t, s, E_0) $ (in red) at various depths $ s $ for one beam, demonstrating the two-dimensional interpolation process on reconstruction pixels (in light grey). 
\label{fig:DDB}}
\end{figure}
%
which has been modeled using the semi-empirical differential Moli\`ere scattering power proposed by Gottschalk \cite{Gottschalk2010, Gottschalk2015}, with the relativistic momentum-velocity product $ pv(s) $ given by the \O{}ver{\aa}s approximation \cite{Overas1960}. The scattering power depends on the material-dependent scattering length $ \chi_s $ and radiation length $ \chi_0 $, which are assumed to take the values of water throughout the entire imaging region. With this assumption, the PB profile is completely characterized as a function of depth.
\subsection{FBP reconstruction with distance-driven binning} \label{Reconstruction}

The reconstruction algorithm utilized is a variation on the PB procedure developed by Rescigno et al. \cite{Rescigno2015}, which utilizes the concept of DDB \cite{Rit2013}. To reconstruct \eqref{eq:Bethe} at some energy $ E_0 $, we have the imaging equation proposed by Wang et al. \cite{Wang2010}, modified for proton PBs using the mean beam paths $ \langle \Lpath \rangle $,
%
\begin{align}
\begin{split}
G_{\theta}(\mu_t, E_0) & \defeq - \int_{E_\mathrm{in}}^{\langle E_\mathrm{out} \rangle} \Bigg[ \frac{S}{\rho} ( I_w, E_0 ) \ \bigg/ \ \frac{S}{\rho} ( I_w, E ) \Bigg] \diff E \\
& \approx \int_{\langle \Lpath \rangle } S \big(\rho_e(\vec{r}\,), I(\vec{r}\,), E_0 \big) \diff \ell, 
\end{split}
\label{eq:Wang}
\end{align}
%
which is weakly-dependent on mean excitation energy $ I $, such that the imaging equation \eqref{eq:Wang} is nearly unaffected by the approximation that $ I $ is set to the value of water, $ I_w $, across all imaged tissues. For all images, we have set $ E_0 \coloneqq E_\mathrm{in} $. The rays $ G $ are calculated using the entrance and average exit energy, at every beam's position-angle touple $ (\mu_t, \theta) $. To form complete projections on the far side of the phantom, a virtual grid of detector pixels is formed for each beam position $ t $,
%
\begin{align}
g(t, s, E_0) &= \frac{ \sum\limits_{i \in B_\theta} h(\mu_i, t, s) G_{\theta}(\mu_i, E_0) }{ \sum\limits_{i \in B_\theta} h(\mu_i, t, s) }, \label{eq:Projections} \\
h(\mu_i, t, s) &= \int_{t - d / 2}^{t + d / 2}  \frac{ \Phi (\mu_i, t, s) }{ N_0 } \ \diff t, \label{eq:Weights}
\end{align}
%
where $ d $ is the length of one pixel. In \eqref{eq:Projections}, the contributions of each beam $ i $ in a projection $ B_\theta $ are combined on a given pixel center at $ t $, according to the weights $ h $ in \eqref{eq:Weights}, which describe the fraction of any bunch's overlap on the pixel.

%
\begin{figure*}[t]
\centering
\subfloat[]{\includegraphics[width=0.24\textwidth]{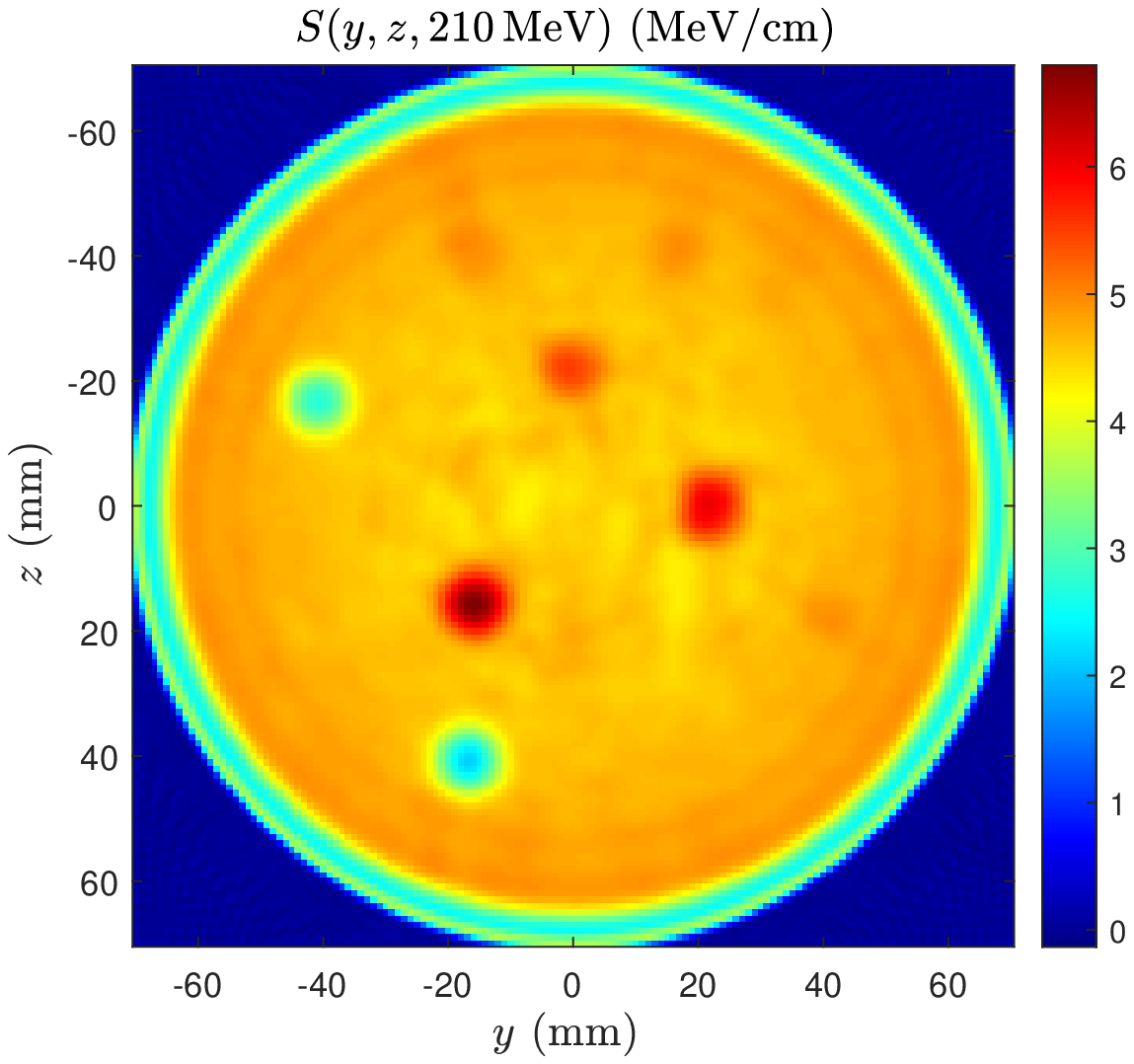}%
\label{fig:StNDdbL}}
\hspace{0.5ex}
\subfloat[]{\includegraphics[width=0.24\textwidth]{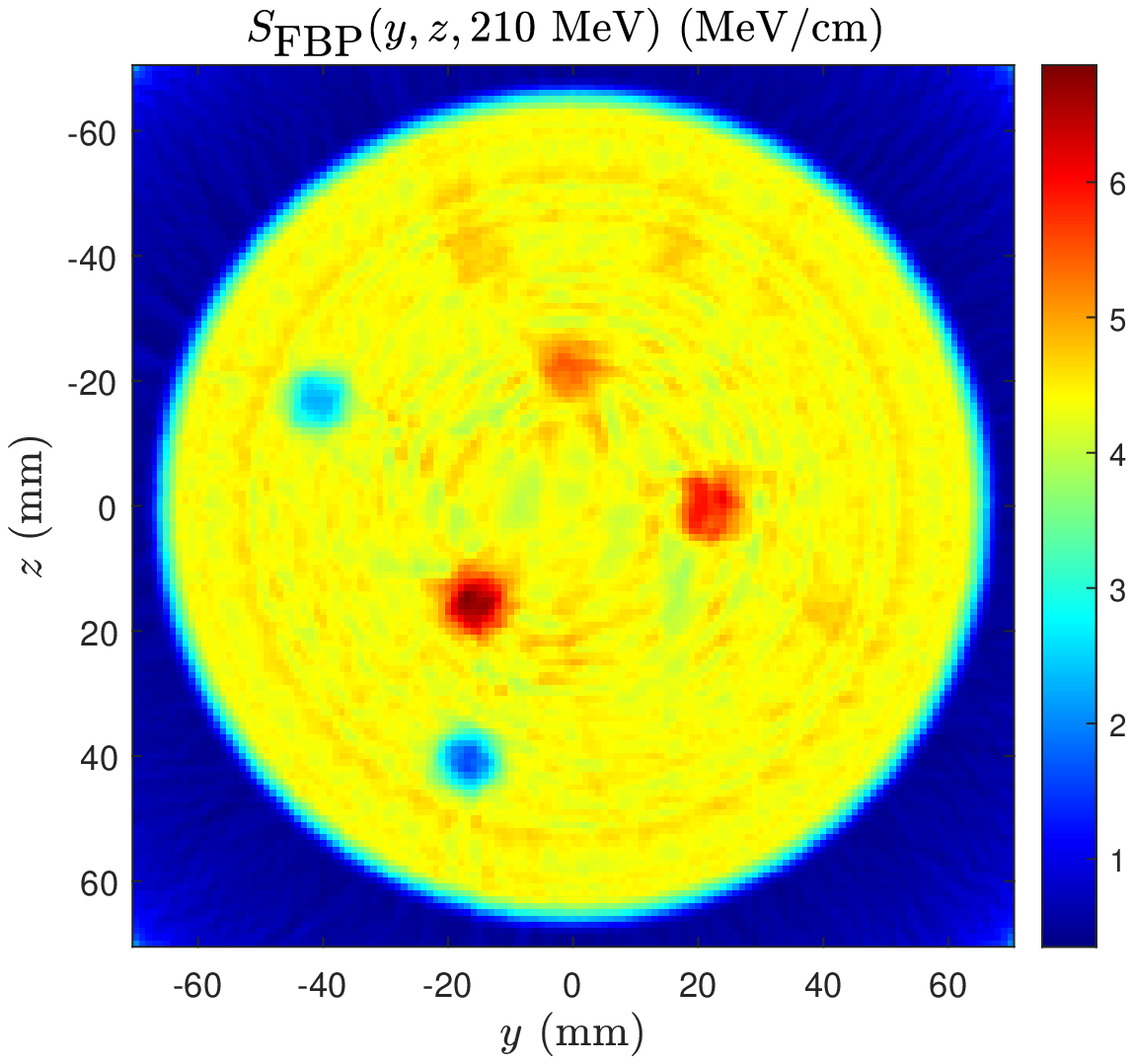}%
\label{fig:StNFbpL}}
\hspace{0.5ex}
\subfloat[]{\includegraphics[width=0.24\textwidth]{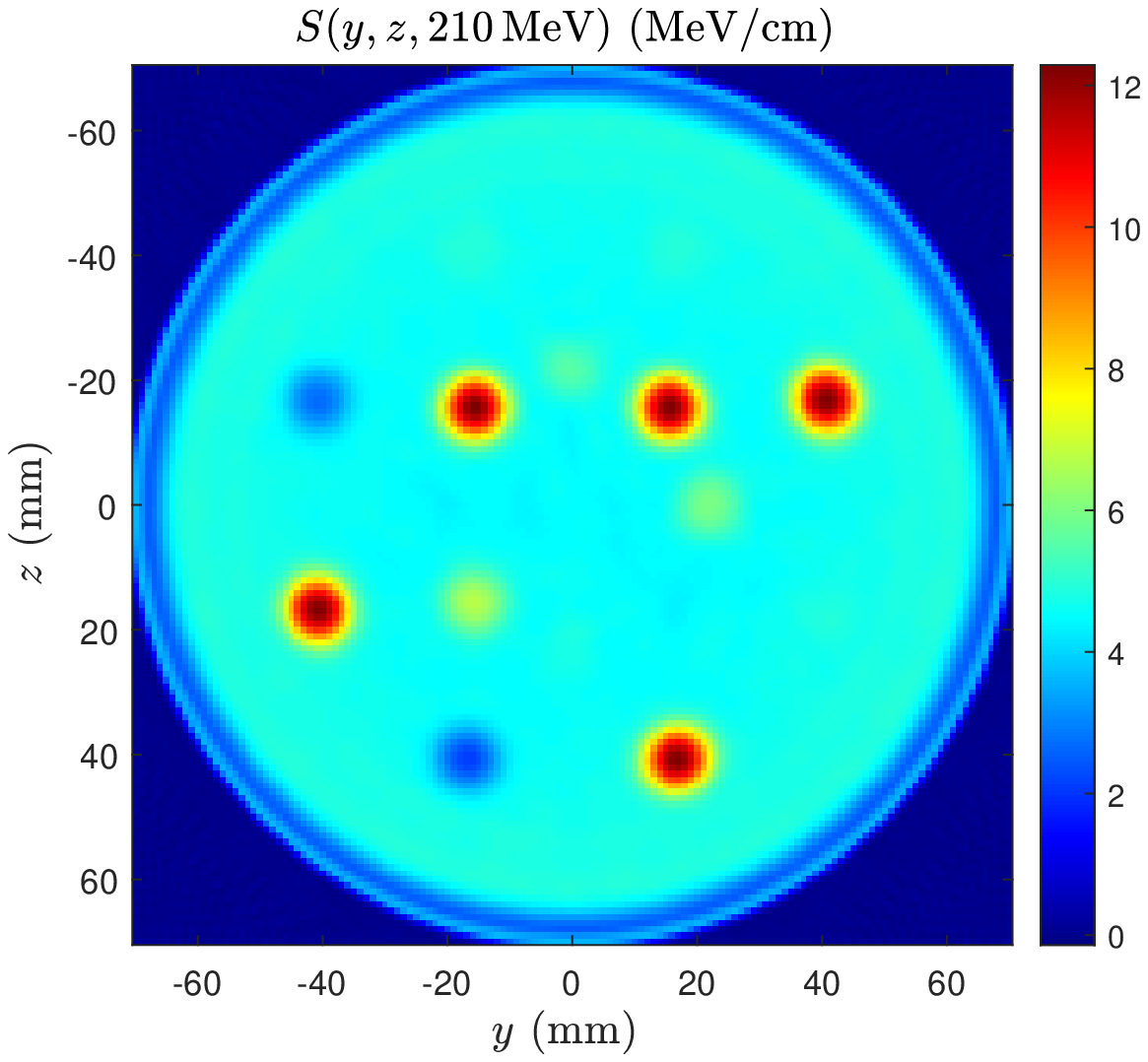}%
\label{fig:TiNDdbL}}
\hspace{0.5ex}
\subfloat[]{\includegraphics[width=0.24\textwidth]{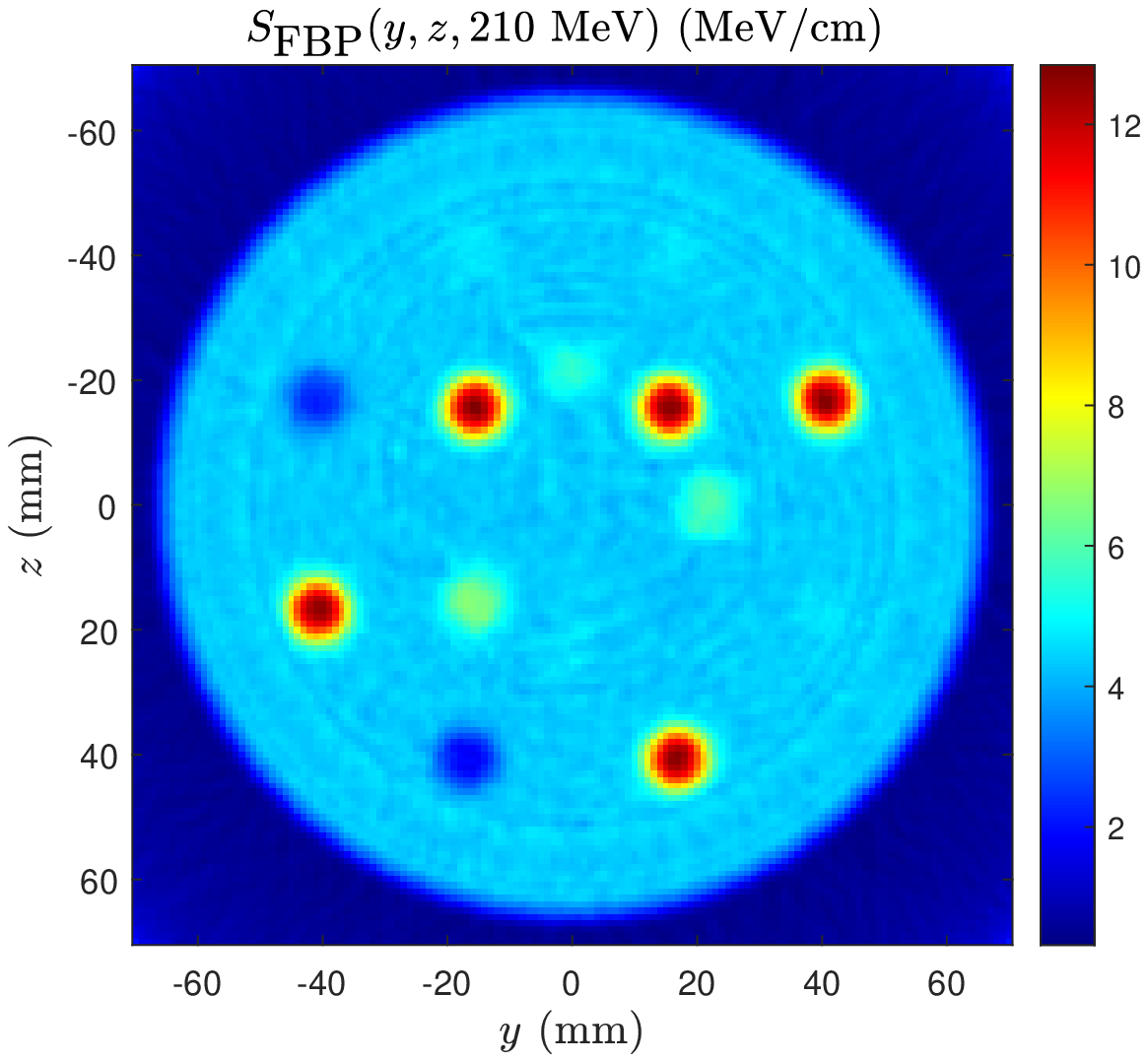}%
\label{fig:TiNFbpL}}
\hfill
\caption{Reconstructed SP images, with a $ \sigma_t(0) = \SI{1.5}{\mm} $ beam, at $ E_\mathrm{in} = \SI{210}{\MeV} $. Images are presented in the following order: (a) standard configuration with a DDB-FBP, (b) standard configuration with a FBP, (c) titanium configuration with a DDB-FBP, and (d) titanium configuration with a FBP. All images are reconstructed with degraded resolution according to the process outlined in Sec. \ref{Images}.
\label{fig:S_Images}}
\end{figure*}
%
%
\begin{figure*}[t]
\centering
\subfloat[]{\includegraphics[width=0.51\textwidth]{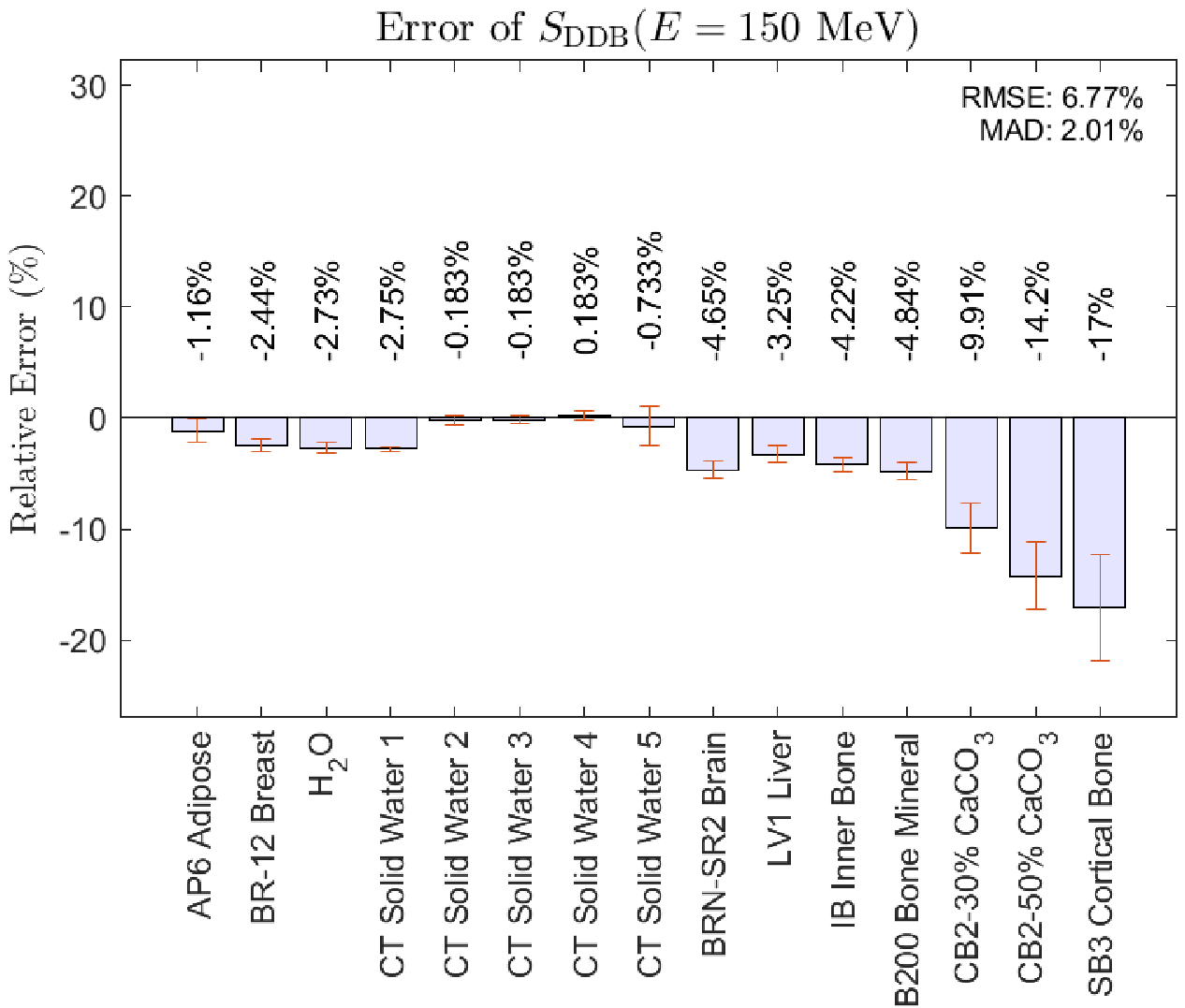}%
\label{fig:StNDdbL_Error}}
\subfloat[]{\includegraphics[width=0.51\textwidth]{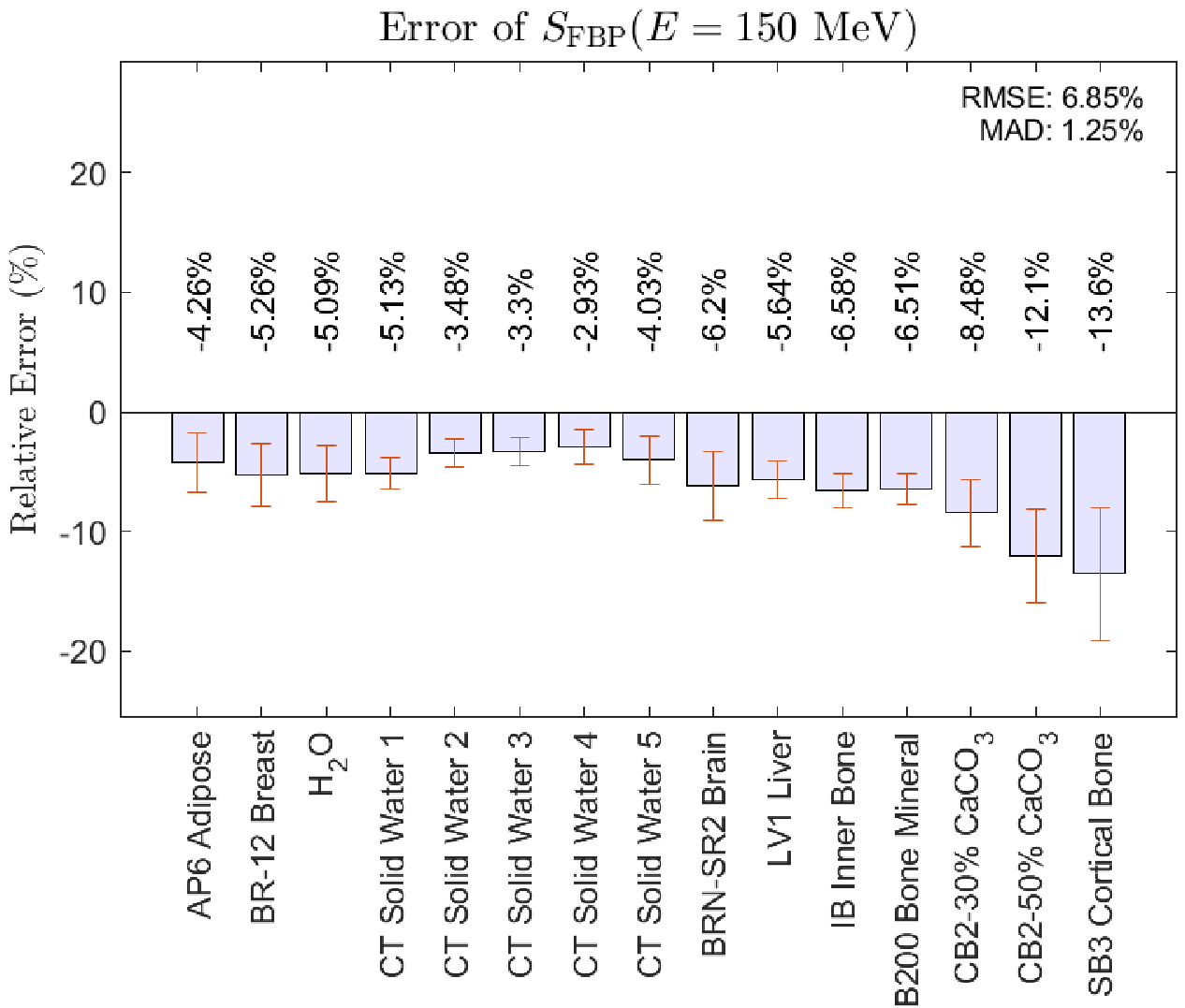}%
\label{fig:StNFbpL_Error}}
\hfill
\caption{Error bars for all soft and bone tissues (listed by increasing mass density) in standard configuration, with $ E_\mathrm{in} = \SI{150}{\MeV} $ and $ \sigma_t (0) = \SI{1.5}{\mm} $. Error bars are one standard error of the mean calculated using all pixels in each insert. Titanium configuration images are omitted for brevity, but show similar profiles.
\label{fig:ErrorBars}}
\end{figure*}
%
The FBP algorithm with DDB (referred to here as the ``DDB-FBP algorithm") utilizes the fact that $ g $ can be evaluated at any depth, rather than just at detector depth. Rather than backprojecting the final beam profile across the entire imaged region, slices of $ g $ are calculated at a number of depths, and backprojected over short distances---where the beam profile is assumed to remain relatively constant. This is done by 2D interpolation at each pixel center $ (t_i,s_i) $ using the grid of calculated $ g(t,s,E_0) $ values. The DDB-FBP algorithm, with reconstruction pixel centers at points $ (y, z) $, is given by 
%
\begin{align}
S(y, z, E_0) \approx \frac{\pi}{N_P} \sum\limits_{i=1}^{N_P} \mathfrak{g}\big( t_i(y,z), s_i(y,z),E_0 \big),
\label{eq:FBPDDB}
\end{align}
%
where the projections $ g $ have been convoluted with a Hamming window to form the filtered projections $ \mathfrak{g} $.
\section{Results} \label{Results}
\subsection{Images and reconstruction parameters} \label{Images}

The number of pixels in the reconstruction grid was set to $ \si{141} \times \si{141} $, and the number of virtual radiographs for DDB was set to \si{200}---no significant increase in accuracy is seen with larger values. For all pixels $ (y, z) $, values of $ \mathfrak{g} $  at $ (t_i(y, z), s_i(y, z)) $ in \eqref{eq:FBPDDB} were interpolated with known nearest-neighbor values using the built-in Matlab function \verb|interp2|. A schematic diagram of this grid of values for a given projection is shown in Fig. \ref{fig:DDB}.

For all data sets reconstructed using the method outlined in Sec. \ref{Reconstruction}, the beam exit energies $ \langle E_\mathrm{out} \rangle $ have been converted into more realistic measurements by inserting them into the empirical calibration curve
%
\begin{align}
s\big( \langle E_\mathrm{out} \rangle \big) = 0.0014 \langle E_\mathrm{out} \rangle^2 + 0.1583 \langle E_\mathrm{out} \rangle - 4.1395,
\label{eq:CalibCurve}
\end{align}
%
after which the resultant depth is rounded to the nearest detector plate center depth (e.g. \SI{0.5}{\mm}, \SI{1.5}{\mm}, etc), and converted back to average exit energy with \eqref{eq:CalibCurve}. In this way, the perfect resolution of the Monte Carlo tallies are degraded to reflect the limitations of the glass calorimeter. For comparison, images with perfect resolution have also been reconstructed without any degradation, to determine the maximum resolution possible utilizing the reconstruction methods presented. These two variations are referred to here as ``perfect resolution" and ``low resolution" images. In addition, a standard FBP algorithm was performed using the Matlab \verb|iradon| function as a benchmark. Both algorithms have been edited to reduce ringing artifacts on the phantom border by applying non-zero padding equal to the end values \cite{Boas2012}, which is applied prior to filtration. Representative images of low-resolution reconstructions at $ E_0 = \SI{210}{\MeV} $ are shown in Fig. \ref{fig:S_Images}.
\subsection{Analysis} \label{Analysis}

To examine the accuracy of the reconstructions, the average value $ \bar{S}_i $ of all pixels geometrically contained within the known boundaries of each insert $ i $ were calculated. Standard errors were calculated from the variance of these pixels according to $ \mathrm{SE}(\bar{S}_i) = \sigma_i \ / \sqrt{N_i} $. Relative errors in Fig. \ref{fig:ErrorBars} are given by
%
\begin{align}
\mathrm{RE}_i = \frac{\bar{S}_i - S_{\mathrm{Ref},i}}{S_{\mathrm{Ref},i}} \times 100 \%,
\label{eq:RelativeError}
\end{align}
%
where reference SP values $ S_{\mathrm{Ref},i} $ are calculated using \eqref{eq:Bethe} with electron densities provided by the manufacturer, and mean ionization energies calculated via Bragg additivity of elemental values taken from Seltzer and Berger \cite{SeltzerBerger1981}. Note that the traditional absolute value bars in \eqref{eq:RelativeError} are omitted to indicate whether the errors are above or below the reference values. The root-mean-square error (RMSE) and mean absolute deviation from the median (MAD) are also displayed for each image, as measures of the image's overall quality. Errors are shown only for standard inserts, although the accuracy of reconstructed soft tissues perform similarly in both phantom configurations.

In addition to relative error, the contrast $ C_i $ of each insert $ i $ is examined according to the following parameter:
%
\begin{align}
C_i = \frac{\left| \bar{S}_i - \bar{S}_{B,i} \right|}{\sqrt{\sigma_i^2 + \sigma_{B,i}^2}}.
\label{eq:Contrast}
\end{align}
%
where $ \sigma_i^2 $ is the variance of the pixels within an insert, $ \bar{S}_{B,i} $ is the average value of approximately $ N_i $ pixels contained within a punctured disk of solid water bordering each insert, and $ \sigma_{B,i}^2 $ the variance of that collection. Contrast for inserts in all eight reconstructions at $ E_0 = \SI{210}{\MeV} $ are shown in Fig. \ref{fig:ContrastPlot}.
\section{Discussion} \label{Discussion}

%
\begin{figure*}[t]
\centering
\subfloat[]{\includegraphics[width=0.47\textwidth]{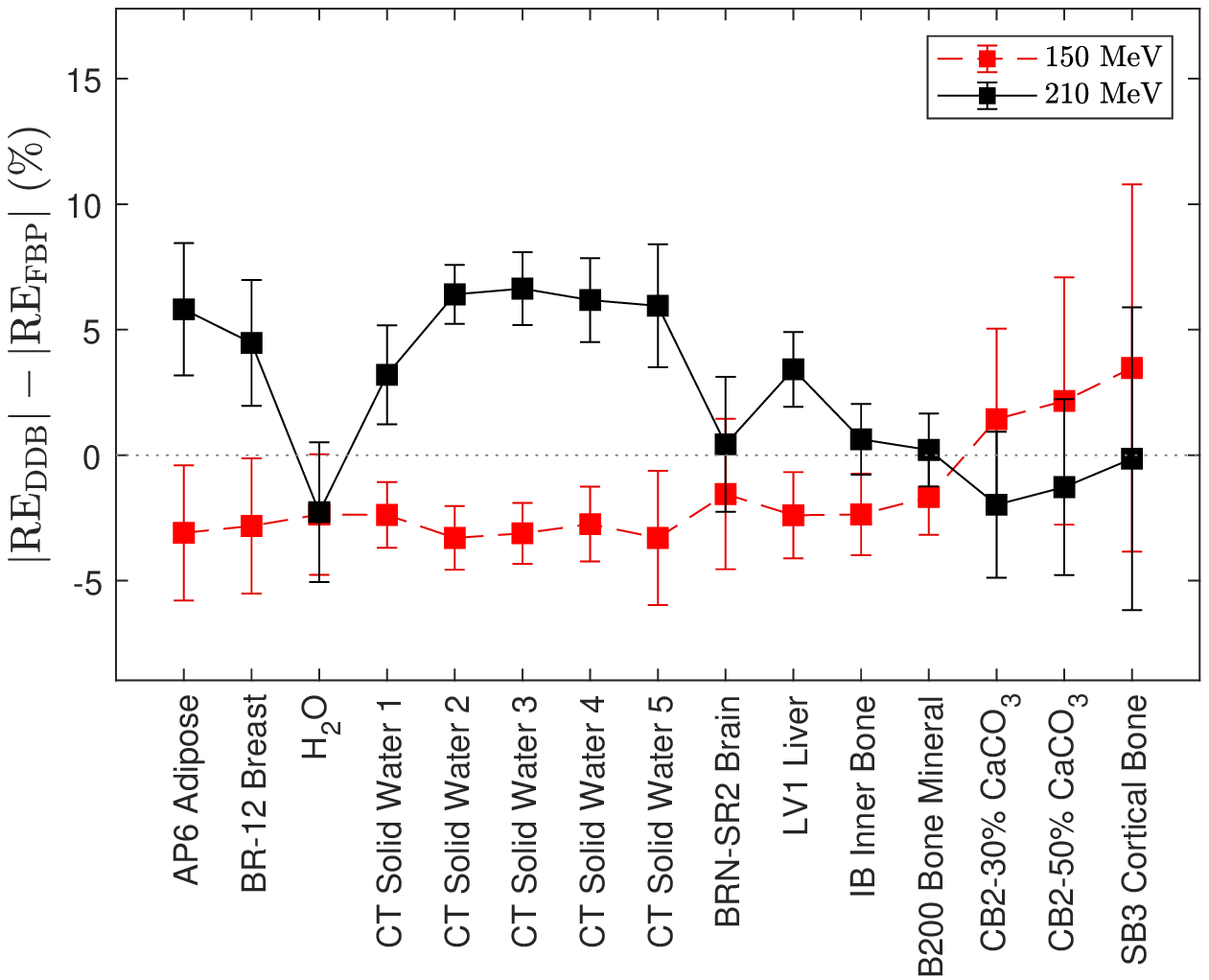}%
\label{fig:FBPvDDB_St}}
\hspace{3ex}
\subfloat[]{\includegraphics[width=0.47\textwidth]{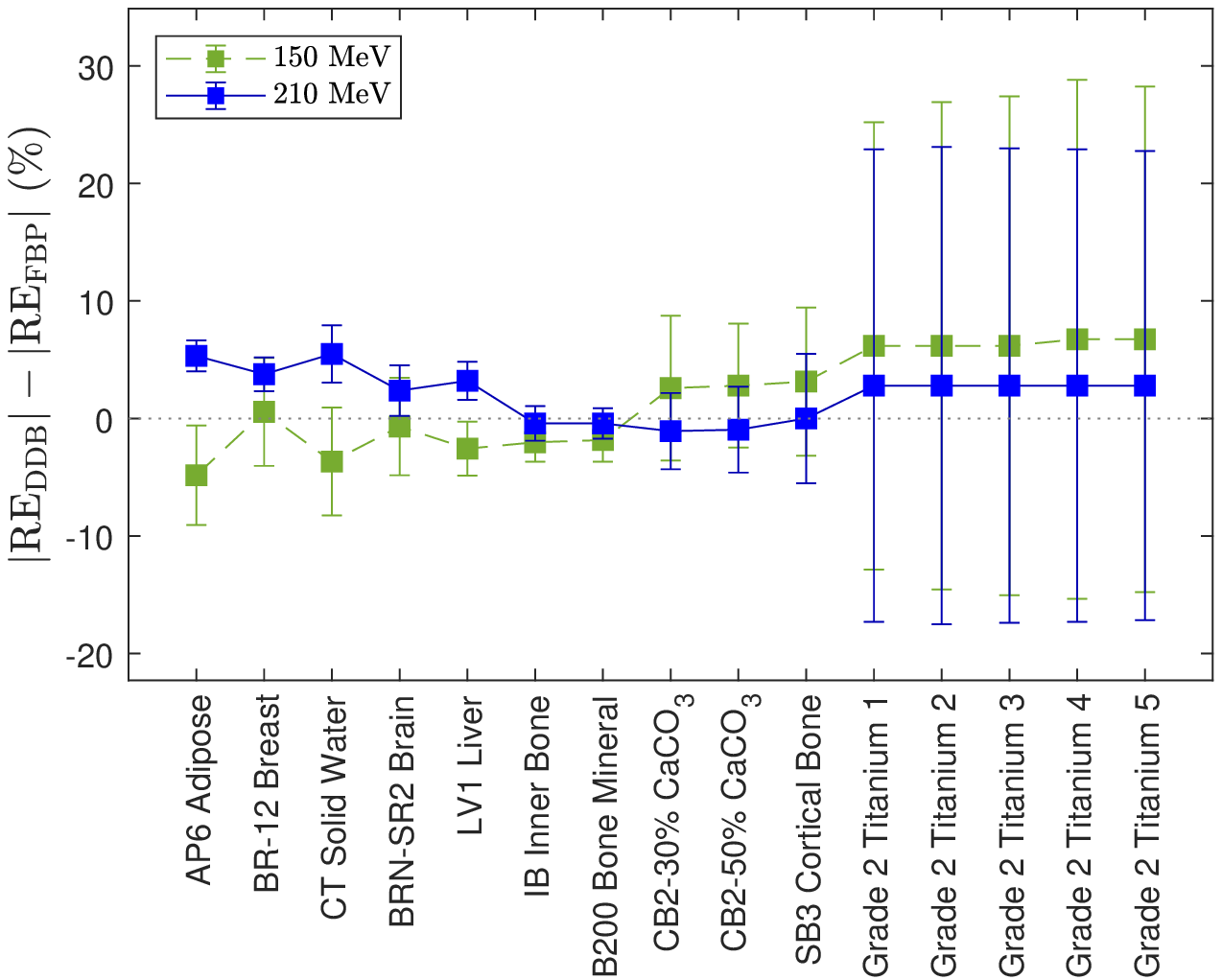}%
\label{fig:FBPvDDB_Ti}}
\hfill
\caption{Comparison of residual errors for the two FBP-style algorithms. Negative values indicate superior performance by the DDB-FBP procedure. Error bars are given by the standard errors of the RE (as in \ref{fig:ErrorBars}) summed in quadrature.
\label{fig:FBPvDDB}}
\end{figure*}
%
Examination of the errors in Fig. \ref{fig:ErrorBars} show relatively small errors for all water-like tissues, with the exception of the image utilizing a low-resolution FBP reconstruction---which is particularly noticeable, as the entire solid water disk background (CT Solid Water 5) carries more than \SI{4}{\percent} error. Only errors for standard configuration reconstructions are shown, at the most well-performing energy simulated (\SI{150}{\MeV})---similar profiles are seen at all energies, but only images reconstructed at $ E_0 = \SI{150}{\MeV} $ show results where the DDB-FBP process arguably outperforms the unaltered FBP. While both reconstruction methods perform similarly with perfect tally resolution, the DDB-FBP algorithm performs somewhat better than the standard FBP for water-like tissues at realistic detector resolution. The errors for lung tissues and titanium (regardless of reconstruction method and resolution) are greater than \SI{20}{\percent} for all images, well outside what could be considered tolerable. All together, these results may demonstrate that a single-proton tracking methodology is necessary to achieve significant results with this detector design. Indeed, the imaging equation of Wang et al. \cite{Wang2010}---with an ART and a YAG:Ce calorimeter---has been tested on a calibration phantom with a similar distribution of tissue substitutes by Civinini et al. \cite{Civinini2020}, with more accurate results.

To examine the DDB-FBP algorithm under a number of different conditions, simulations were carried out with the four aforementioned energies. A pattern is seen in Fig. \ref{fig:FBPvDDB}: regardless of the distribution of inserts in the phantom, the DDB-FBP algorithm performs well at low-density (less than calcium carbonate) and low-energy. At higher densities, distance-driven binning is favorable at higher energy, but is still outperformed by the standard FBP. Fig. \ref{fig:FBPvDDB} plots only \SI{150}{\MeV} and \SI{210}{\MeV} reconstructions for visual clarity, but \SI{170}{\MeV} and \SI{190}{\MeV} data lie between the two in a natural pattern. A larger, $ \sigma_t(0) = \SI{2.5}{\mm} $ beam was also simulated at \SI{210}{\MeV} to mimic the spread of the IBA dedicated nozzle \cite{Moignier2016}, to examine the performance of a wider beam. The stopping powers reconstructed with this data set are generally of low quality, and DDB does not appear to improve upon the FBP algorithm, despite it accounting for beam spread by design. The respectable performance of unaltered FBP at narrow beam widths is not particularly surprising, as the standard deviation of the beam increases by less than \SI{1.09}{\mm} after traversing \SI{19.9}{\cm} of water at $ E_\mathrm{in} = \SI{210}{\MeV} $, making the straight-line assumption of rays in CT backprojection fairly reasonable. The results of these alterations in beam energy and spread show that the addition of DDB is of limited use, and requires specific conditions to improve upon the unmodified FBP. Results may be improved by altering the DDB-FBP procedure to account for the spread of material-dependent $ \chi_s $ and $ \chi_0 $ parameters across the phantom.

\begin{figure*}[t]
\centering
\subfloat[]{\includegraphics[width=0.45\textwidth]{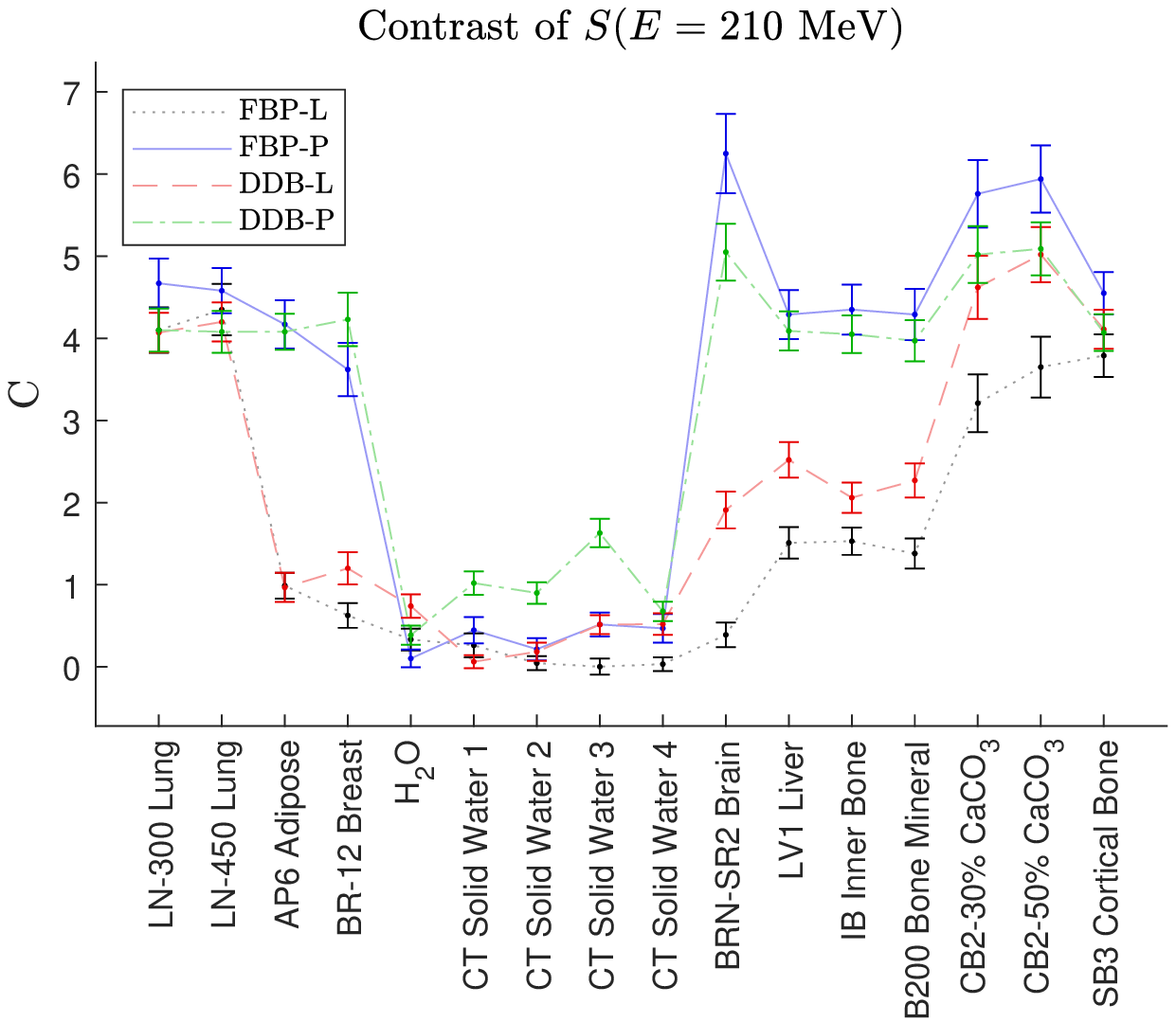}%
\label{fig:St_Contrast}}
\hspace{3ex}
\subfloat[]{\includegraphics[width=0.45\textwidth]{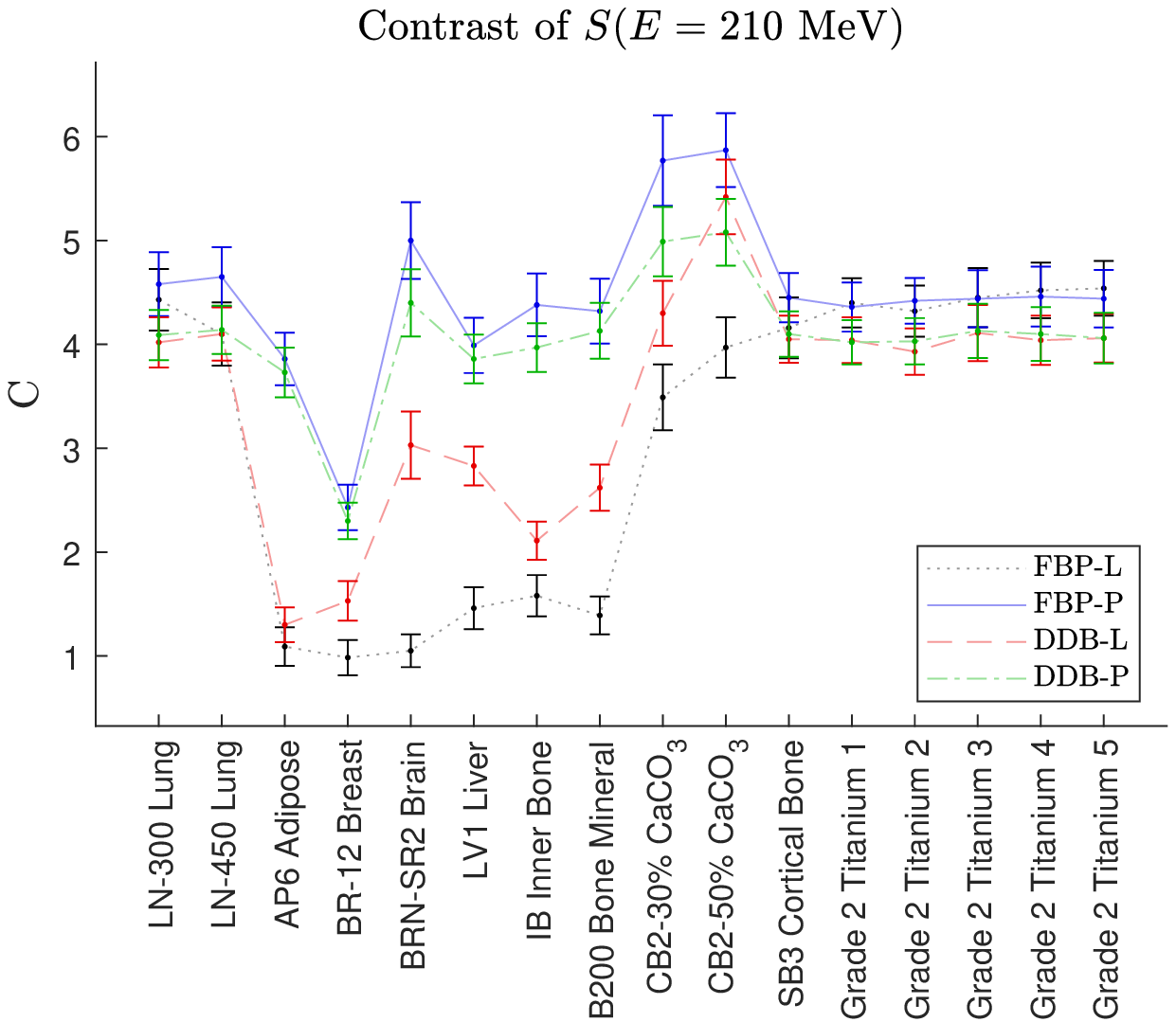}%
\label{fig:Ti_Contrast}}
\hfill
\cprotect\caption{Contrast for all eight images at $ E_0 = \SI{210}{\MeV} $ with $ \sigma_t(0) = \SI{1.5}{\mm} $. All confidence intervals are given by one standard error of \eqref{eq:Contrast} using \si{10000} bootstrap samples for each insert with the Matlab internal function: \verb|bootstrp|. ``L" and ``P" in the legends refer to the degraded and un-degraded resolutions referenced in Sec. \ref{Images}.
\label{fig:ContrastPlot}}
\end{figure*}

Contrast values for all inserts are shown in Fig. \ref{fig:ContrastPlot}. Both algorithms perform similarly, though those created with DDB-FBP have slightly more visible inserts in the vicinity of $ \rho = \SI{1}{\g\per\cm\cubed} $ in the titanium configuration. Brain, liver, adipose, and breast tissue substitutes are not visible using either reconstruction at low resolution in standard configuration, but slightly increased contrast is seen using DDB. Similar contrast profiles are seen at all reconstructed energies, and only images at $ E_0 = \SI{210}{\MeV} $ are shown. Noisy artifacts are more visible using the unmodified FBP at low resolution, making most water-like tissues nearly indistinguishable from the background. Less pronounced streaks are also seen using the DDB-FBP, indicating that more projections may be required for accurate reconstructions.

It should also be noted that the boundaries between insert and solid water disk appear somewhat subjective---especially for low- and high-density inserts---despite \eqref{eq:RelativeError} and \eqref{eq:Contrast} averaging over exactly all pixels calculated to be within the known boundaries of the MCNP simulation. These theoretical insert boundaries are unlikely to be respected by the reconstructed images, but any large variation in values within an insert will be reflected in the error bars of Fig. \ref{fig:ErrorBars}.
%
%

Certain shortcomings of the simulated glass must also be addressed. As described by Taki et al. \cite{Taki2013}, a high $ \ce{WO4} $-content in the proposed system tends to promote the formation of opaque $ \ce{\alpha-Gd2(WO4)3} $ crystals within the glass network. Consequently, the glass simulated in this study represents a best case scenario, and slightly lower density glasses presented by Tillman et al. \cite{Tillman2017} may need to be considered. Alternatively, higher density glasses in the europium-doped tungsten gadolinium borate system have been reported, achieving densities as high as \SI{6.173}{\g\per\cm\cubed} with increased europium dopant \cite{Wantana2020}.
\section{Conclusion} \label{Conclusion}

Monte Carlo simulations of pCT acquisitions have been carried out to test the feasibility of FBP-style algorithms used with a compact glass calorimeter and proton therapy beams. Simulations were performed for two different calibration phantom configurations, with various different initial beam conditions to test the algorithm over a wide range of possible beam properties. Reconstructed images appear qualitatively accurate at thin beam widths, and the DDB-FBP algorithm performs well at the low end of therapeutic proton beam energies. However, these images are likely not yet quantitatively accurate enough to justify the use of proton imaging with the glass detector over x-ray CT stoichiometric methods---or more modern dual-energy CT to stopping power conversions procedures. This indicates that more accurate algebraic reconstruction methods may improve upon the results given here, and are likely necessary to utilize this detector design to its fullest potential. With a more suitable reconstruction algorithm, the proposed glass calorimeter may be a valuable addition to existing therapy gantries.

\bibliography{ms}
\bibliographystyle{IEEEtran}

\vfill

\end{document}